\documentclass[pre, showpacs, aps, superscriptaddress]{revtex4}
\usepackage{amsfonts}
\usepackage{epsfig}

\newcommand{\w}{\omega}
\newcommand{\pd}[2]{\frac{\partial #1}{\partial #2}}
\newcommand{\dd}[2]{\frac{d #1}{d #2}}
\renewcommand{\k}{{\bf k}}
\newcommand{\kerKE}[1]{{\mathcal F}_4\left[ #1\right]}
\newcommand{\ptwod}[2]{\frac{\partial^2 #1}{\partial #2 ^2}}
\newcommand{\ws}{{\omega^*}}
\newcommand{\ts}{{t^*}}
\newcommand{\ddd}[2]{\frac{d^2 #1}{d #2 ^2}}
\newcommand{\dddd}[2]{\frac{d^3 #1}{d #2 ^3}}
\newcommand{\ddddd}[2]{\frac{d^4 #1}{d #2 ^4}}
\newcommand{\Dt}{\Delta t}
\newcommand{\K}{{\mathcal K}}

\newcommand{\Kn}{{\mathcal K}\left[n_\w\right]}
\newcommand{\pdddd}[2]{\frac{\partial^4 #1}{\partial #2 ^4}}
\newcommand{\pddd}[2]{\frac{\partial^3 #1}{\partial #2 ^3}}
\newcommand{\pdd}[2]{\frac{\partial^2 #1}{\partial #2 ^2}}

\begin{document}

%\begin{frontmatter}
\title{Non-stationary Spectra of Local Wave Turbulence}
\author{Colm Connaughton}
\affiliation{Mathematics Institute, University of Warwick,
Coventry CV4 7AL, U.K.}
\author{Alan C. Newell}
\affiliation{Department of Mathematics, University of Arizona,
Tucson, AZ 85721, U.S.A.}
\affiliation{Mathematics Institute,
University of Warwick, Coventry CV4 7AL, U.K.}
\author{Yves Pomeau}
\affiliation{Laboratoire de Physique Statistique de l'Ecole
Normale Sup\'erieur, associ\'e au CNRS, 24 Rue Lhomond, 75231
Paris Cedex 05, France}
\affiliation{Department of Mathematics,
University of Arizona, Tucson, AZ 85721, U.S.A.}

%\begin{abstract}
%Abstract for the Physica D front paper

\begin{abstract}
The evolution of the Kolmogorov-Zakharov (K-Z) spectrum of weak
turbulence is studied in the limit of strongly local interactions
where the usual kinetic equation, describing the time evolution of
the spectral wave-action density, can be approximated by a PDE. If
the wave action is initially compactly supported in frequency
space, it is then redistributed by resonant interactions producing
the usual direct and inverse cascades, leading to the formation of
the K-Z spectra. The emphasis here is on the direct cascade. The
evolution proceeds by the formation of a self-similar front which
propagates to the right leaving a quasi-stationary state in its
wake. This front is sharp in the sense that the solution remains
compactly supported until it reaches infinity. If the energy
spectrum has infinite capacity, the front takes infinite time to
reach infinite frequency and leaves the K-Z spectrum in its wake.
On the other hand, if the energy spectrum has finite capacity, the
front reaches infinity within a finite time, $\ts$, and the wake
is steeper than the K-Z spectrum. For this case, the K-Z spectrum
is set up from the right after the front reaches infinity. The
slope of the solution in the wake can be related to the speed of
propagation of the front. It is shown that the anomalous slope in
the finite capacity case corresponds to the unique front speed
which ensures that the front tip contains a finite amount of
energy as the connection to infinity is made. We also introduce,
for the first time, the notion of entropy production in wave
turbulence and show how it evolves as the system approaches the
stationary K-Z spectrum.
\end{abstract}

%\end{abstract}

\maketitle
%\end{frontmatter}

%Intro to Physica D front paper
\section{Introduction and Motivation}
\label{sec-intro} Wave turbulence is concerned with the
statistical description of an infinite sea of dispersive waves,
which are weakly coupled by nonlinear interactions and maintained
away from equilibrium by interaction with sources and sinks of
energy. The theory has found practical application in many
branches of physics including the description of surface waves on
fluid
interfaces\cite{hasselmann1962,hasselmann1963,benney67,zakharov66},
Alfven wave turbulence in astrophysical plasmas
\cite{galtier2000,ng96}, nonlinear optics\cite{dyachenko92} and
acoustics\cite{newell71,zakharov70} to name a few.

The central quantity of theoretical interest is the spectral wave
action density, $n_\k$, which describes how the excitations in the
system are distributed among different wave-vectors, $\k$. Under
fairly weak assumptions\cite{newell2001}, the long time behaviour
of $n_\k$ is given by an equation known as the wave kinetic
equation. For a system dominated by four wave interactions this
equation takes the form,
\begin{equation}
\label{eq-4WKE_k} \pd{n_\k}{t}= 4\pi \int
\left|T_{\k\k_1\k_2\k_3}\right|^{\ 2} \ \kerKE{n_\k}
\delta(\k\!+\!\k_1\!-\!\k_2\!-\!\k_3)\ d\k_1 d\k_2 d\k_3,
\end{equation}
where
\begin{equation}
\hspace*{-0.5in}\label{eq-4WKE-F} \kerKE{n_\k} =n_\k
n_{\k_1}n_{\k_2}n_{\k_3}
\left(\frac{1}{n_\k}+\frac{1}{n_{\k_1}}-\frac{1}{n_{\k_2}}
-\frac{1}{n_{\k_3}}\right)
\delta(\omega_{\k}\!+\!\omega_{\k_1}\!-\!\omega_{\k_2}\!-\!\omega_{\k_3}).
\end{equation}
Equation (\ref{eq-4WKE_k}) is the analogue for waves of the
Boltzmann equation of classical statistical mechanics. In many
applications, $\w_\k$ and $T_{\k\k_1\k_2\k_3}$ are homogeneous
functions of their arguments. Their degrees of homogeneity shall
be denoted by $\alpha$ and $\gamma$ respectively. Under rescaling,
$\k \to \lambda\k$, they transform as follows :
\begin{eqnarray}
\label{eq-homogeneityDegrees}
 \w_{\lambda\k} =\lambda^\alpha\w_\k\\
 T_{\lambda\k\lambda\k_1\lambda\k_2\lambda\k_3}&=&\lambda^\gamma
T_{\k\k_1\k_2\k_3}.
\end{eqnarray}
It was shown by Zakharov\cite{zakharov} in the 60's that if the
energy sources and sinks are separated by an ``inertial range'',
equation (\ref{eq-4WKE_k}) has exact isotropic steady state
solutions,
\begin{eqnarray}
\label{eq-KZ1}n_\k &=& c_1P^{1/3}k^{-(2\gamma+3d)/3}\\
\label{eq-KZ2}n_\k &=& c_2Q^{1/3}k^{-(2\gamma+3d - 1)/3},
\end{eqnarray}
which carry constant fluxes of conserved densities,  in this case
energy flux, $P$, or wave action flux, $Q$, between sources and
sinks. These steady state spectra are the direct analogues of the
direct and inverse cascades in hydrodynamic turbulence and are
referred to as Kolmogorov-Zakharov (K-Z) spectra. They have been
well observed experimentally in a variety of contexts.

 In 1991 Falkovich and Shafarenko\cite{falkovich91}
addressed the question of how the K-Z spectrum is set up in time
if $n_\k$ is initially compactly supported in wave-vector space.
They used a self-similar solution of (\ref{eq-4WKE_k}) and an
assumption that, for the direct cascade, the total energy
increases linearly in time, to show that (\ref{eq-KZ1}) is set up
by a nonlinear front which propagates towards $k=\infty$ and
leaves the $k^{-(2\gamma+3d)/3}$ spectrum in its wake.

However subsequent numerical simulations of the kinetic equation
for Alfven wave turbulence performed by Galtier et al.
\cite{galtier2000} suggested that the development of the K-Z
spectrum may proceed by a different route. For the Alfven wave
system, the K-Z energy spectrum has finite energy capacity,
meaning that on the K-Z spectrum, $\int E(\k)\ d\k <\infty$. This
implies that the nonlinear front must reach $k=\infty$ within a
finite time, $\ts$. They noticed that the spectrum in the wake of
the front was significantly steeper than the K-Z value for times
less than the singular time, $\ts$, and that the K-Z spectrum then
developed from right to left after the front reached $k=\infty$.
Other work by Pomeau et al.\cite{pomeau2001} on the inverse
cascade in the Nonlinear Schrodinger equation suggested that there
might be anomalous quasi-stationary spectra associated with
non-stationary solutions of kinetic equations. However no-one has
yet made a specific attempt to search for them.

One of the challenges in far-from-equilibrium systems is to
understand the means by which stationary states are reached and to
ask if there are functionals analogous to the entropy in
equilibrium systems. What we will show is that while the entropy,
which for wave turbulence is formally (see for example \cite{aucoin1971})
\begin{equation}
S=\int \ln n_\k\,d\k,
\end{equation}
is not well defined on the steady state solutions, spectra
(\ref{eq-KZ1}) and (\ref{eq-KZ2}), its production rate is. We find
that for $0 < t < \ts$, when the spectrum in the wake of the front
is steeper than the K-Z spectrum, the entropy production is
positive. At $\ts$ the connection to $\k=\infty$ is made and
energy is no longer a conserved quantity. For $t > \ts$, the K-Z
spectrum is established via a front which travels back from $\k
=\infty$. During this stage, the entropy production rate, while
still positive, gradually decreases and asymptotes to zero, its
value on the exact K-Z spectrum. We conjecture that this scenario,
established in this paper for the differential approximation to
the kinetic equation, (\ref{eq-4WKE_k}), will be widely valid for
finite capacity non-equilibrium systems, including
three-dimensional hydrodynamic turbulence at large Reynolds
numbers.

This leads us to the topic of this article. We have made an
extensive study of the non-stationary solutions of the so-called
differential kinetic equation of local wave turbulence. This model
equation is obtained from (\ref{eq-4WKE_k}) under the assumption
that the interaction co-efficient, $T_{\k\k_1\k_2\k_3}$ is
strongly localised in $\k\k_1\k_2\k_3$ space. It has the advantage
of replacing the integro-differential kinetic equation with a PDE.
We find that the qualitative bahaviour observed by Galtier et al
is present in this model in the finite capacity case. Since we are
dealing with a PDE, we can go a lot further in terms of
understanding.

The organisation of the article is as follows. In section
\ref{sec-dke} we introduce the differential kinetic equation and
describe a few of its properties which make it a good model of
wave turbulence. We also introduce exact expressions  for the
fluxes of energy ($P$), the flux of particles ($Q$), and the
entropy production rate in terms of its flux ($R$) and its bulk
production rate ($T$). The latter is always positive definite. We
calculate each of these quantities on the algebraic solutions,
$n_\k \propto k^{-\alpha x}$.  Section \ref{sec-pde_num} contains
the details of some numerical simulations of the PDE. These
simulations suggest that the nonlinear front is ``sharp'' in the
sense that $n_\k$ remains compactly supported for $t<\ts$. There
is a singularity, a divergence in the second derivative in fact,
at the front tip between the regions $n_\k=0$ and $n_\k>0$. Next,
in section \ref{sec-ss}, we construct a family of self-similar
solutions of the differential kinetic equation which are
parameterised by a single free parameter. This free parameter can
be interpreted as the asymptotic slope behind the front. Following
that, in section \ref{sec-margin}, we use this self-similarity
analysis to formulate a hypothesis which we call the {\it critical
front speed hypothesis}.  This hypothesis is based on physical
arguments and allows us to select a critical value, $x_c$, for the
asymptotic slope, given by
\begin{equation}
\label{eq-magic_formula} x_c=x_0+\frac{2\gamma-3\alpha}{12\alpha},
\end{equation}
where $x_0$ denotes the usual K-Z exponent for the direct cascade.
This formula is well supported by our numerical simulations. In
our conclusion, we attempt to make a connection with entropy
production arguments. Two appendices are provided. In the first,
we analyse the mathematical structure of the similarity equation
and try to understand how the critical slope is related to the
solution trajectories of the underlying ordinary differential
equation. The second appendix contains a brief outline of the
numerical methods used.

% Review of results and notation relating to the DKE
\section{The Differential Kinetic Equation}
\label{sec-dke}

We begin by briefly discussing the origin of the differential
kinetic equation. Assuming that the wave-action spectrum rapidly
becomes isotropic, and averaging over angles, we can make a
transformation from $d$-dimensional wave-vector space to frequency
space,
\begin{equation}
\label{eq-4WKE-w} \pd{N_\w}{t}= \int S_{\w\w_1\w_2\w_3}\
\kerKE{n_\w}\ d\w_1 d\w_2 d\w_3
\end{equation}
where
\begin{equation}
\hspace*{-0.5in}\label{eq-interaction-w} S_{\w\w_1\w_2\w_3} =
4\pi\!\int \!\!\left|T_{kk_1k_2k_3}\right|^{\ 2}
\delta(\k\!+\!\k_1\!-\!\k_2\!-\!\k_3)\
(kk_1k_2k_3)^{d-1}\dd{k}{\w}\dd{k_1}{\w_1}\dd{k_2}{\w_2}\dd{k_3}{\w_3}
\ d\Omega
\end{equation}
and $N_\w$ is defined by requiring that
\begin{equation}
\int \phi(\w) N_\w d\w=\int \phi(\left|\k\right|^\alpha)\ n_\k
d\k,
\end{equation}
for any test function, $\phi$.
 Here the
volume element $d\Omega$ represents integration over the angular
variables in $\k_1\k_2\k_3$ space and the wave-vector moduli are
related to the frequency via the dispersion relation,
\begin{equation}
\label{eq-isotropicDispersion} \w_k = ck^\alpha.
\end{equation}
If we assume that the interaction coefficient,
$T_{\k\k_1\k_2\k_3}$ is strongly local in $\k\k_1\k_2\k_3$ space
then (\ref{eq-4WKE-w}) can be approximated by a differential
equation \cite{dyachenko92},
\begin{equation}
\label{eq-dke} \pd{N_\w}{t} = I\ptwod{ }{\w}\left(
\w^{s}n_\w^4\ptwod{ }{\w}\left(\frac{1}{n_\w} \right)\right),
\end{equation}
where
\begin{eqnarray}
n_\w(t)&=&n(k(\w),t)\\
s&=&3x_0+2\\
x_0&=&\frac{2\gamma+3d}{3\alpha}.
\end{eqnarray}
$I$ is a pure number which comes from the angular integrations in
(\ref{eq-interaction-w}). This equation is called the differential
kinetic equation.

The local approximation leading to the differential kinetic
equation, (\ref{eq-dke}), is rather drastic and is not justified
for many of the physical applications of weak turbulence.
Nonetheless, it retains many of the qualitative features of the
full kinetic equation, (\ref{eq-4WKE-w}). It provides an excellent
model in the context of which these features can be easily
studied. In particular, the differential kinetic equation respects
the conservation laws embodied within its integro-differential
precursor, namely the conservation of the total energy,
\begin{equation}
 E=\int \w N_\w\ d\w,
\end{equation}
and the total number of particles,
\begin{equation}
N=\int  N_\w\ d\w.
\end{equation}
It also preserves the scaling and homogeneity properties of the
kinetic equation. Consequently, the pure scaling solutions of the
kinetic equation, the equilibrium thermodynamic spectra and the
non-equilibrium K-Z spectra, are also solutions of the
differential kinetic equation. These solutions are of the form
$n_\w=c\w^{-x}$ where $x$ takes one of the following values:
\begin{displaymath}
x=1,\hspace{0.5cm} x=0,\hspace{1.0cm}\mbox{\rm (thermodynamic)}
\end{displaymath}
or
\begin{displaymath}
x=\frac{2\gamma+3d}{3\alpha},\hspace{0.5cm}
x=\frac{2\gamma+3d-\alpha}{3\alpha},\hspace{1.0cm}\mbox{\rm
(K-Z).}
\end{displaymath}
It is convenient to define
\begin{equation}
\label{eq-DKE_K} \Kn = \w^{3x_0+2}n_{\w}^4\ptwod{
}{\w}\left(\frac{1}{n_\w}\right).
\end{equation}
The two conservation laws can be written as continuity equations :
\begin{eqnarray}
\label{eq-DKE_E_cons}\pd{N_\w}{t} - \pd{Q}{\w}&=&0\\
\label{eq-DKE_N_cons}\pd{E_\w}{t} + \pd{P}{\w}&=&0,
\end{eqnarray}
where
\begin{equation}
\label{eq-DKE_Q} Q=\pd{\K}{\w}
\end{equation}
is the local flux of particles,
\begin{equation}
\label{eq-DKE_P} P=\K-\w\pd{\K}{\w}
\end{equation}
is the local flux of energy and
\begin{equation}
\label{eq-DKE_E} E_\w=\w N_\w
\end{equation}
is the energy density if frequency space.
 Note that $P$ is defined to be
positive when energy flows to the right in $\w$ space and $Q$ is
positive when particles flow to the left.

\begin{figure}
\begin{center}
\epsfig{file=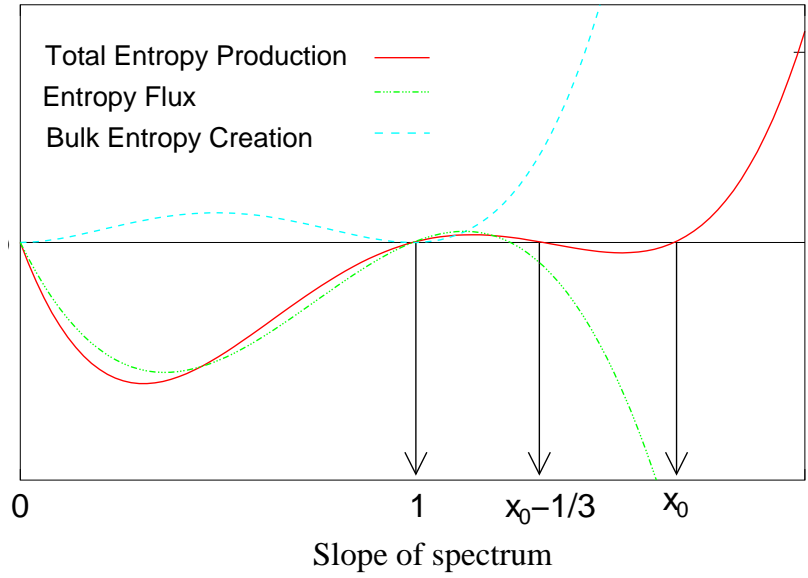,height=3.0in,angle=0}
\end{center}
\caption{\label{fig-entropy} Plot of the bulk entropy production,
$T$, the entropy flux, $R$, and the total entropy production,
$\dot{S}$, as a function of the exponent, $x$, of the spectrum for
power law spectra, $n_\w=c\w^{-x}$. }
\end{figure}

The entropy, $S$, of a wave system is formally $\int \ln
n_\k\,d\k$. The production rate,
\begin{equation}
\dd{S}{t}=\dd{ }{t}\,\int \ln n_\k\,d\k = \int
\frac{1}{n_\w}\,\dd{N_\w}{t}\,d\w,
\end{equation}
can readily be calculated to be
\begin{equation}
\dot{S} = -\pd{R}{\w} + T.
\end{equation}
The entropy flux, $R$, which is positive for entropy flow to large
wave-numbers, is
\begin{equation}
R=-\frac{Q}{n_\w^2}\pd{ }{\w}(\w n_\w) -
\frac{P}{n_\w^2}\pd{n_\w}{\w}
\end{equation}
and the local bulk entropy production rate, $T$, is
\begin{equation}
T = I \w^{3x_0+2}n_\w^4\left(\ptwod{
}{\w}\left(\frac{1}{n_\w}\right)\right)^2.
\end{equation}
Note that $T$ is positive definite and zero on the thermodynamic
solution, $n_\w=\tau/(\w-\mu)$ , $\tau$ being the temperature and
$\mu$ the chemical potential. Indeed if the system were isolated,
say on the interval $\w_1<\w<\w_2$, and $P$, $Q$ and $R$ were
identically zero, then the usual principles of equilibrium systems
would apply.

On the solution $n_\w=c\w^{-x}$, the quantities $K$, $Q$, $P$,
$R$, $T$ and $-\pd{R}{\w}+T$ are calculated. They are:
\begin{eqnarray}
K&=& I\,c^3\w^{3x_0-3x}x(x-1)\\
Q&=& 3Ic^3x(x-1)(x_0-x)\,\w^{3x_0-3x-1}\\
P&=& 3Ic^3x(x-1)(x-(x_0-{\scriptstyle
\frac{1}{3}}))\,\w^{3x_0-3x}\\
T &=& c^2\,I\,x^2(x-1)^2\,\w^{3x_0-2x-2}\\
R &=& 3 c^2\,I\,x(x-1)(x_0-{\scriptstyle\frac{4}{3}}x)\,\w^{3x_0-2x-1},\\
\label{eq-localEP}\dot{S} &=& -\pd{R}{\w} + T=
9c^2\,I\,x(x-1)(x-x_0)(x-(x_0-{\scriptstyle
\frac{1}{3}}))\,\w^{3x_0-2x-2}.
\end{eqnarray}
Similar expressions, which have the {\em same} zeros (as functions
of $x$) obtain for the case when the differential approximation is
replaced by the full collision integral. In particular, we note
that the entropy production rate is always positive (we assume
$x_0>4/3$) for $1<x<x_0-1/3$ and for $x>x_0$, the K-Z exponents
for particles and energy respectively. The relevant functions are
plotted in figure \ref{fig-entropy}. For $x_0-1/3 < x < x_0$, the
entropy production rate is negative. This corresponds to a
situation when the particle flux is building a condensate state
\cite {pomeau2001}.

%Presentation of the numerical results on the PDE version of the DKE
\section{Numerical Observations of Non-stationary Spectra}
\label{sec-pde_num}

Let us now turn our attention to non-stationary solutions of
(\ref{eq-dke}). In particular we are interested in how the K-Z
spectra are set up if we begin from an initial condition which is
compactly supported at low frequencies. Early work on this
question focussing on the direct cascade by Falkovich and
Shafarenko \cite{falkovich91} suggested that the K-Z spectrum is
set up by a nonlinear front which propagates to the right leaving
the K-Z spectrum in its wake. Recent numerical studies by Galtier
et al \cite{galtier2000} suggest that this problem is more subtle.
They found that in the case where the K-Z energy spectrum has
finite capacity, the approach to the steady state proceeds by a
different mechanism. For this system the nonlinear front reaches
infinite frequency within a finite time, $t^*$. They found that
the quasi-stationary spectrum in the wake of this front was
actually steeper than the K-Z spectrum. The K-Z spectrum was then
set up from right to left after the front reached infinity.
\begin{figure}
\begin{center}
\epsfig{file=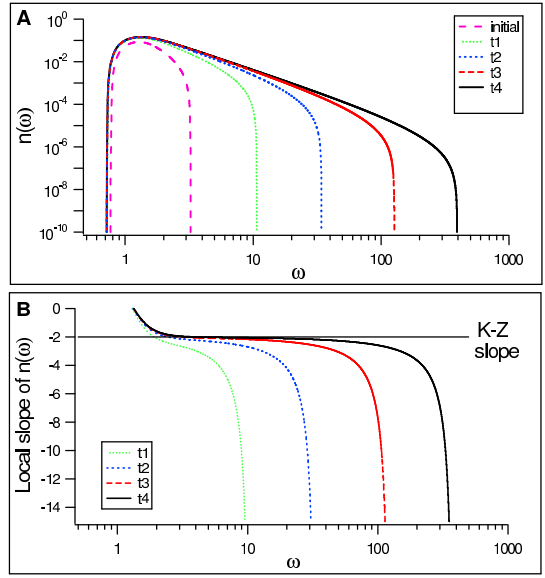,height=3.0in,angle=0}
\end{center}
\caption{\label{fig-evol_inf}Time evolution of the spectrum and
the local slope for the parameter values  $\alpha=0.5, \gamma=0,
d=1$. The logarithmic scales are to base 10.}
\end{figure}

\begin{figure}
\begin{center}
\epsfig{file=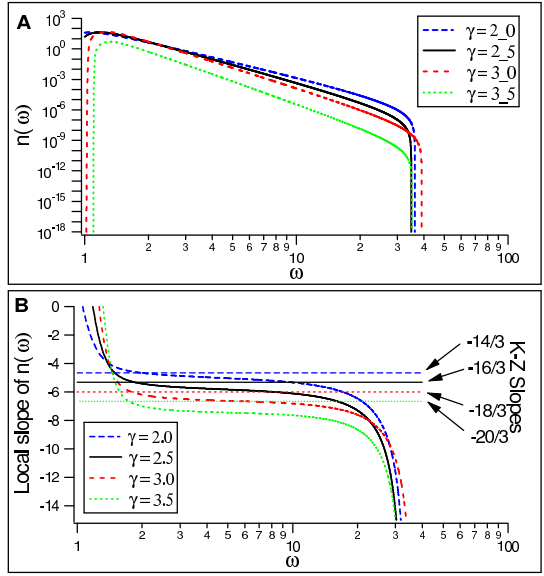,height=3.0in,angle=0}
\end{center}
\caption{\label{fig-evol_fin}Snapshots of the non-stationary
spectrum for a range of different values of the parameter
$\gamma$, keeping $\alpha=1/2$ and $d=1$. The logarithmic scales
are to base 10.}
\end{figure}

We investigated whether there was evidence of this behaviour in
the solutions of the differential kinetic equation. We solved the
differential kinetic equation numerically and followed the
evolution from an initial condition compactly supported at low
frequencies. The results presented in this section were obtained
by allowing the initial data to decay freely in the absence of
forcing or damping. Some details of the numerical methods used are
contained in appendix \ref{app-numerics}.

Let us first consider the situation in which the energy spectrum
has infinite capacity, $2\gamma < 3\alpha$. Figure
\ref{fig-evol_inf} shows a sequence of snapshots of $n_\w$ and the
local slope at successive times for a system with parameter values
$\alpha=0.5, \gamma=0, d=1$. We conclude that the slope tends to
the K-Z value far behind the front in agreement with previous
expectations. The K-Z spectrum is set up asymptotically in time
from the left of the window of transparency to the right.

Now consider what happens when we increase the value of $\gamma$
and bring the system into the finite capacity regime. Figure
\ref{fig-evol_fin} shows the absolute value of the local slope of
$n_\w$ at similar stages in the evolution for a sequence of values
of $\gamma$ in the finite capacity regime.  $\gamma$ takes the
values  2.0, 2.5, 3.0 and 3.5. Visually, it is clear that as
$\gamma$ increases, the slope of the spectrum behind the front
tends to a value which is increasingly steeper than the K-Z value.
This observation is supported by fitting power law functions to
the numerical data. The ``best fit'' slopes are presented in Table
\ref{tab-anomalous_slopes} along with the K-Z values for
comparison.  The slope given by equation (\ref{eq-magic_formula})
agrees well with the numerical observations. The time, $t^*$,
required for the front to reach infinity depends both on the
parameters $\gamma$ and $\alpha$ and on the initial energy
distribution. We have not, in the present work, made any
systematic attempt to understand this aspect of the problem.

\begin{table}
  \centering
  \begin{tabular}{|c|c|c|c|}
    \hline
    $\gamma$& $x_{KZ}$& x$_{\rm num}$ & x$_{c}$\\
    \hline
    2.0 & -4.67 & -5.12 & -5.08 \\
    2.5 & -5.33 & -5.93 & -5.92 \\
    3.0 & -6.00 & -6.67 & -6.75 \\
    3.5 & -6.67 & -7.50 & -7.58 \\ \hline
  \end{tabular}
  \caption{Numerical evidence for the
  anomaly. The table shows the values of the wake slopes obtained by
  fitting numerical data for a range of values of $\gamma$ in the finite
  capacity regime. These are to be compared to the K-Z values and the
  values of the critical slope obtained from equation (\ref{eq-magic_formula}) :
  $x_c = x_{KZ}+(2\gamma-3\alpha)/12\alpha$.
  \label{tab-anomalous_slopes}}
\end{table}

\begin{figure}
\begin{center}
\epsfig{file=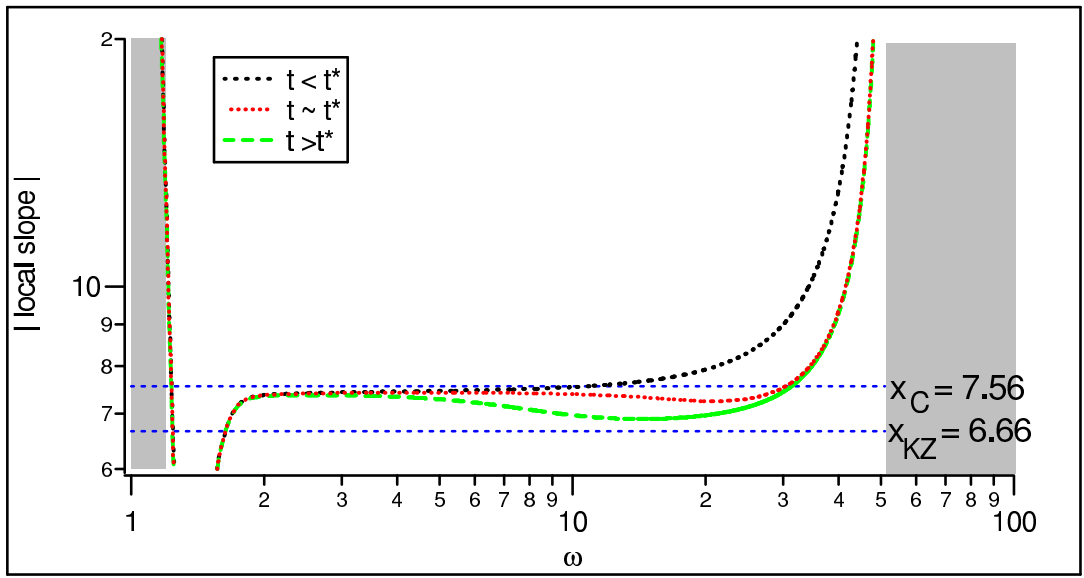,height=3.0in,angle=0}
\end{center}
\caption{\label{fig-relax}Log-log plot showing the initial stages
of relaxation to the K-Z slope after the front reaches $\w_R$.
Parameter values are $\gamma=7/2$, $\alpha=1/2$ and $d=1$. The
absolute value of the local slope is plotted as a function of
frequency for three different times, just before $\ts$,
approximately at $\ts$ and after $\ts$. There is a clear
transition from the steeper slope, $x_c$, to the shallower
$x_{KZ}$.}
\end{figure}
The next obvious question  is that of how the system makes the
transition from this quasi-steady anomalous regime to the K-Z
spectrum which we know to be the final steady state. For finite
capacity systems this transition begins once the front reaches
infinity. Clearly we cannot easily treat the divergence of the
front numerically. Instead we allow the front to propagate into a
regime of strong damping to the right of the window of
transparency. This damping region is intended to mimic the energy
sink role provided by the point $\w=\infty$. Figure
\ref{fig-relax} shows the local slope of $n_\w$ for successive
times after the front reaches the dissipation region. The results
of Figure \ref{fig-relax} are for $\gamma=3.5$ where there is a
significant difference of about 0.9 between the anomalous slope
and the K-Z slope. We see that the system begins to relax towards
the K-Z spectrum as soon as the dissipation scale is reached.
Notice that the relaxation is occurring from right to left.

Another issue which is crucial to our explanation of the finite
capacity anomaly is the structure of the front itself. From our
numerical simulations we found that the front tip seems to be
``sharp'' in the sense that the spectral wave action density
remains on compact support during the time evolution. Such sharp
fronts have been known to exist in solutions of certain classes of
nonlinear diffusion equations dating back to the work of
Zel'dovich in the 50's. \cite{zeldovich50} Subsequent work by
Lacey et al. \cite{lacey82} showed that such fronts can be
stationary, moving or exhibit waiting time behaviour where the tip
remains stationary for a finite time before beginning to move.

\begin{figure}
\begin{center}
\epsfig{file=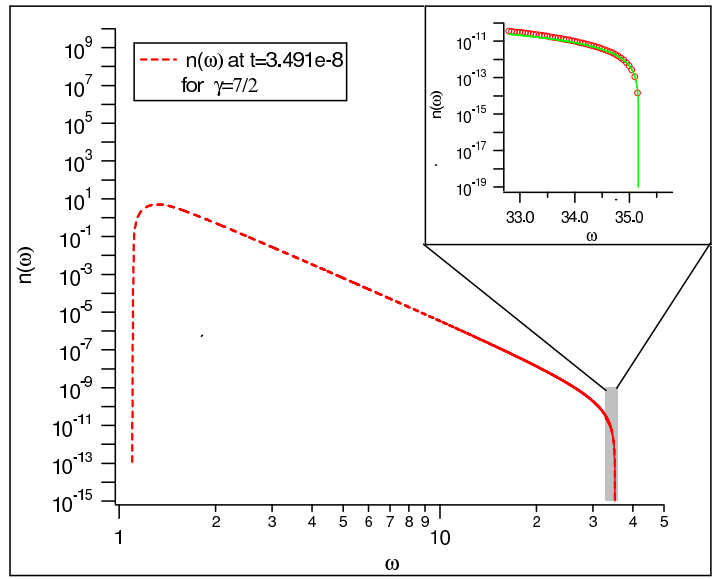,height=3.0in,angle=0}
\end{center}
\caption{\label{fig-front_tip}Inset shows a detailed view of the
structure of the front tip seen in the evolution of a finite
capacity spectrum. The parameter values are $\gamma=7/2$,
$\alpha=1/2$ and $d=1$. The line is a least squares fit of the
numerical data to the functional form $n(\w) \sim A(\ws-\w)^c$ as
suggested by equation (\ref{eq-front_tip}). The fitted value of
$c$ is 1.52.}
\end{figure}

Let us suppose that our system possesses a sharp tip  located at $\ws(t)$ and
\begin{displaymath}
n(\w,t) \left\{
\begin{array}{ll}
=0& \hspace{0.5cm}\mbox{for $\w\geq\ws(t)$}\\
>0&\hspace{0.5cm}\mbox{for $\w<\ws(t)$.}\end{array}
\right.
\end{displaymath}
To the left of $\ws(t)$, we look for a solution of equation
(\ref{eq-dke}) in the form of a power series,
\begin{equation}
\label{eq-front_tip_series} n(\w,t)=A(t)\sum_i
a_i(\ws(t)-\w)^{m_i},
\end{equation}
where the constants $a_i$ and the exponents $m_i$ are to be
determined by expanding equation (\ref{eq-dke}) around $\w=\ws(t)$
and substituting this representation for $n(\w,t)$. The leading
order term on the LHS is
\begin{equation}
A\ws^{\frac{d}{\alpha}-1}m_0a_0(\ws-\w)^{m_0-1}\dd{\ws}{t}.
\end{equation}
The leading order term on the RHS is
\begin{equation}
A^3\ws^sm_0(m_0+1)(3m_0-2)(3m_0-3)a_0^3(\ws-\w)^{3m_0-4}.
\end{equation}
Comparing powers of $\ws-\w$ immediately yields $m_0=3/2$. We
cannot fix the time dependence although we get the following
relation between $A(t)$ and $\ws(t)$,
\begin{equation}
\label{eq-front_tip_time_dependence} \ws(t)^{\frac{d}{\alpha}-1}
\dd{\ws}{t} = A^2(t) \ws(t)^s,
\end{equation}
after choosing $a_0^{-2}=(m_0+1)(3m_0-2)(3m_0-3)$. The presence of
a sharp tip and the structure
\begin{equation}
\label{eq-front_tip} n(\w,t) \sim A(t)a_0(\ws(t)-\w)^{3/2},
\end{equation}
immediately behind the tip is well supported by our numerical
simulations as shown in figure \ref{fig-front_tip}.

% Construction of self similar solutions
\section{Self-Similar Solutions of the Differential Kinetic
Equation} \label{sec-ss}

To study non-stationary solutions of (\ref{eq-dke}) analytically,
make the following self-similarity ansatz for the form of the
solution,
\begin{equation}
\label{eq-ss_solution} n_\w(t) = \ws(t)^{-x}\ F(\eta),
\end{equation}
where the self-similar variable, $\eta$, is defined by
\begin{equation}
\label{eq-ss_var} \eta=\frac{\w}{\ws(t)}.
\end{equation}
Under this change of variables the derivatives transform according
to the relations,
\begin{eqnarray}
\label{eq-ss_change_of_vars} \pd{ }{t} &=& \dot{\ws}\left(\pd{
}{\ws} - \eta\ws^{-1}\pd{ }{\eta}\right)\\
\pd{ }{\w} &=& \ws^{-1}\pd{ }{\eta},
\end{eqnarray}
and equation (\ref{eq-dke}) can be rewritten as
\begin{equation}
\label{eq-ss_dke} A\
\eta^{\frac{d}{\alpha}-1}\left(-xF-\eta\dd{F}{\eta}\right)= \ddd{
}{\eta}\left(\eta^sF^4\ddd{
}{\eta}\left(\frac{1}{F}\right)\right),
\end{equation}
where
\begin{equation}
\label{eq-ss_cond}
 A=\dot{\ws}\ \ws^{2x+d/\alpha - 3x_0}.
\end{equation}
Here $x_0=(2\gamma+3d)/3\alpha$ is the exponent of the K-Z energy
spectrum. The free parameter, $x$, is the asymptotic slope of the
spectrum far behind the front. This is because we assume that
there exists a quasi-stationary regime far behind the front, which
is a simple power law to leading order, $n_w\sim\w^{-y}$. We must
therefore have $y=x$ in order to cancel the time dependence from
the leading order part of (\ref{eq-ss_solution}).

For a self-similar solution, $A$ must be time independent. We are
interested in situations where the system has finite energy
capacity and generates a singularity within a finite time which we
shall denote by $\ts$. The appropriate solution of
(\ref{eq-ss_cond}) describing such situations is
\begin{equation}
\label{eq-ws} \ws(t)=(\ts-t)^b.
\end{equation}
It is convenient to define
\begin{equation}
\label{eq-kappa_d}\kappa_d=\frac{2\gamma-3\alpha}{3\alpha}.
\end{equation}
The direct cascade has infinite energy capacity for $\kappa_d<0$
and finite energy capacity for $\kappa_d>0$. Upon substitution of
the form (\ref{eq-ws}) into (\ref{eq-ss_cond}) it follows that
\begin{eqnarray}
\label{eq-b}b&=&\left(2(x-x_0)-\kappa_d\right)^{-1}\\
\label{eq-S}A&=&-b
\end{eqnarray}
We can also define
\begin{equation}
\label{eq-kappa_i}\kappa_i=\frac{2\gamma-\alpha}{3\alpha},
\end{equation}
such that the inverse cascade has infinite particle capacity for
$\kappa_i>0$ and finite particle capacity for $\kappa_i<0$.
Equation (\ref{eq-b}) can be written in the equivalent form
\begin{equation}
\label{eq-b2}b=\left(2(x-x_0+1/3)-\kappa_i\right)^{-1}
\end{equation}
which is more appropriate for studying the inverse cascade. Notice
that the speed of propagation of the front  at $\ws$, as  measured
by $b$, is related to the asymptotic slope, $x$, behind the front.

If we are interested in the direct cascade then $\ws\to\infty$ as
$t\to\ts$. This corresponds to $b<0$. Conversely, for the inverse
cascade, $\ws\to 0$ as $t\to\ts$, corresponding to $b>0$.

Upon substitution of (\ref{eq-b}) and (\ref{eq-S}) into
(\ref{eq-ss_dke}) we obtain
\begin{equation}
\label{eq-ss_dke2} (2(x-x_0)-\kappa_d)^{-1}\
\eta^{-1+d/\alpha}\left(xF+\eta\dd{F}{\eta}\right)= \ddd{
}{\eta}\left(\eta^sF^4\ddd{
}{\eta}\left(\frac{1}{F}\right)\right).
\end{equation}

Let us now consider the structure of the front tip in the
self-similar variables. From our numerical simulations we expect
that there is a singularity in the solution as $\eta \to 1$. For
the direct cascade, we look for  a singularity of the form
$(1-\eta)^m$, approaching from the left. For the inverse cascade
we expect a singularity of the form $(\eta-1)^m$ approaching from
the right. Let us restrict our attention to the direct cascade.

Consider an expansion of the solution to the left of $\eta=1$ in
the form
\begin{equation}
\label{eq-near_right_singularity} F(\eta) =
(1-\eta)^m\sum_{n=0}^{\infty} a_n(1-\eta)^n
\end{equation}
Taylor expand equation (\ref{eq-ss_dke}) to the left of $\eta=1$
and substitute this expansion for $F$. Matching of the leading
order divergences fixes $m=3/2$. In principle the coefficients,
$a_n$, can be computed to arbitrary order by matching powers of
$1-\eta$. The first few of them, computed using {\it Mathematica},
are
\begin{eqnarray}
a_0&=&\frac{2}{5}\,\sqrt{\frac{2}{3}}\,{\sqrt{-b}}\\
a_1&=&\frac{{\sqrt{-b}}}{27\,\sqrt{6}\,\alpha} \,\left( -3\,d + 7\,\alpha\,s - 2\,\alpha\,x \right)  \\
\nonumber a_2&=& \frac{\sqrt{-b}}{1075032\,{\sqrt{6}}\,\alpha^2}\,
       \left(18405\,d^2 -6\,\alpha\,d\,\left( 4860 + 7931\,s - 3490\,x \right)\right.\\
       &&  + \left.\alpha^2\,\left( 47117\,s^2 -20\,x\,\left( 1944 + 359\,x \right)
       -28\,s\,\left( -4374 + 833\,x \right)  \right)\right)
\end{eqnarray}

\begin{figure}
\begin{center}
\epsfig{file=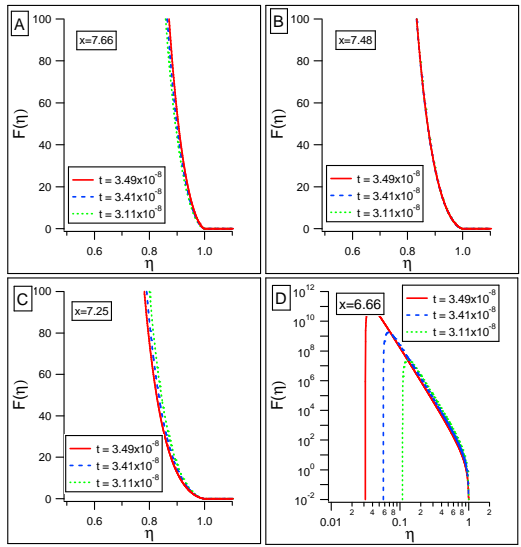,height=3.5in,angle=0}
\end{center}
\caption{\label{fig-ss_data} Application of the similarity
transformation, (\ref{eq-ss_solution}), to the direct cascade at
three consecutive times We take a range of values for the
parameter $x$. Graph A takes $x=7.66$, graph B takes $x=7.48$,
graph C takes $x=7.25$. The self-similarity of the direct cascade
and the front tip is evident from graph B.  For reference, graph D
takes $x=6.66$, corresponding to K-Z scaling, and  uses a log plot
to display the entire function $F(\eta)$. It is clearly not
self-similar for this value of $x$. }
\end{figure}

So far, the slope behind the front, $x$, or equivalently by
(\ref{eq-b}), the front speed, $b$, is a free parameter in the
similarity transformation. We wish to understand how the system
picks the particular value for this parameter. If we are to
observe the Kolmogorov-Zakharov slope behind the front then b must
have the value
\begin{equation}
\label{eq-b_KZ} b_{KZ} =  -1/\kappa_d,
\end{equation}
which follows from putting $x=x_0$ in equation (\ref{eq-b}). In
section \ref{sec-pde_num} we presented numerical evidence that the
slope behind the front was anomalous. We can also confirm this
anomaly and check the correctness of the self-similarity argument
outlined in this section by applying the transformations
(\ref{eq-ss_solution}), (\ref{eq-ss_var}) to the solution of the
PDE. In figure \ref{fig-ss_data} we have taken a particular
solution of (\ref{eq-dke}) at a number of successive times and
applied the similarity transformation for a selection of values of
the free parameter $x$. The data presented is for
the case $\alpha=1/2$, $d=1$ and $\gamma=7/2$. It is clear that
the direct cascade is self-similar for $x=7.48$ approximately.
This is to be compared with the theoretical prediction $x_c=7.56$.
The corresponding transformation with K-Z value, $x=6.66$ in this
case, is clearly not self-similar.

The behaviour of the solution of the equation for the spectrum
after the singular time, $\ts$, is an interesting question. Let us
consider what could be called the ``bouncing back'' of the
spectrum from infinity after $t=\ts$. Our considerations are
inspired by \cite{pomeau2001}. Exactly at $t=\ts$ the spectrum
becomes a pure power law, $\w^{-x_c}$, at large frequencies. This
can be understood as follows: any finite value of $\w$, however
large, is in the wake part of the self-similar solution when
$t=\ts$. This wake is a pure power spectrum. Therefore at $t=\ts$,
we have a well defined initial condition for the evolution
equation.

This spectrum is not a stationary solution (neither K-Z nor
equilibrium) of the evolution equation. The subsequent evolution
should follow the same principles as just before $t=\ts$: the
large frequency part of the spectrum has a typical timescale which
goes to zero as $\omega \to \infty$. This timescale is the
timescale for relaxation to a stationary K-Z spectrum with
constant energy flux. Although the amplitude of this spectrum
itself changes in the course of time, it does so more and more
slowly after $t=\ts$ so that the changes induced in the K-Z
spectrum by the change of energy flux become adiabatic relative to
the infinitely short timescale for the large frequency evolution.
This justifies our consideration of the large time spectrum as a
K-Z spectrum although it is not strictly speaking stationary in
time.

Let assume now that the bouncing back of the K-Z spectrum from
infinity is described by the same self-similar equation as before
$t=\ts$. One boundary condition is now different. The support of
the self-similar spectrum now goes from zero to infinity and as
$\w\to\infty$, $n_\w \sim \w^{-x_0}$ with $x_0$ being the K-Z
exponent. Near small frequencies on this stretched scale, the
spectrum keeps the same power behaviour as before $t=\ts$ since
the timescale there is long compared to the timescale $\ts-t$ at
the high frequency end. We expect that arguments which we shall
present in section \ref{sec-ode} are relevant here too, with the
corresponding solution of the self-similar equation being unique
and the free parameter required to adjust the trajectory reaching
the low frequency behaviour along the stable manifold being the
amplitude of the K-Z spectrum at infinity. The solution of the
self-similar equation should then specify the shape of the
``bend'' in the spectrum seen in figure \ref{fig-relax} and its
time evolution will be then given be the same scaling of the
frequencies in the similarity representation of the dynamical
equations as we used before $t=\ts$. Some further work will be
required to make these statements more concrete.

%Careful derivation and discussion of the marginality argument

\section{Derivation of the Anomalous Spectrum from the Critical Front Speed Hypothesis}
\label{sec-margin}

In this section we present a heuristic derivation of the formula
(\ref{eq-magic_formula}) promised in the introduction. The program
is as follows. We first calculate the energy balance in the
neighbourhood of the front tip. We will find that the condition
that energy is conserved by the motion of the tip is equivalent to
the condition, (\ref{eq-ss_cond}), that the solution be self
similar. Next we calculate the amount of energy, $E(t_1,t_2)$,
which enters the the frequency interval
$\left[\ws(t_1),\ws(t_2)\right]$ immediately behind the tip in the
time interval $\left[t_1,t_2\right]$. The critical front speed
hypothesis is the following : {\it
 the physical system selects the
unique value of the front speed such that $\lim _{t_2\to \ts}
E(t_1,t_2)$ is finite and nonzero for all $t_1< \ts$.}

The basic energy balance equation for the front tip is as follows
\begin{equation}
\label{eq-energy_balance1} \int_{t_1}^{t_2}
\int_{\ws(t_1)}^{\ws(t_2)} \left(\pd{E}{t}+\pd{P}{\w}\right)\ d\w\
dt = 0
\end{equation}
This can be expanded to read
\begin{equation}
\int_{\ws(t_1)}^{\ws(t_2)}\left(E(\w,t_2)-E(\w,t_1)\right)\ d\w =
-\int_{t_1}^{t_2}\left(P(\ws(t_2),t)-P(\ws(t_1),t) \right)\ dt
\end{equation}
We observe that $E(\w,t_1)=0$ since $\ws(t_1)<\w$ and
$P(\ws(t_2),t)=0$ since $\ws(t)<\ws(t_2)$. Hence we obtain the
fundamental balance equation for the front tip,
\begin{equation}
\label{eq-energy_balance2} \int_{\ws(t_1)}^{\ws(t_2)} E(\w,t_2) \
d\w =\int_{t_1}^{t_2} P(\ws(t_1),t)\ dt.
\end{equation}

Let us now calculate $P$ and $E$ near the front tip using the
expansion (\ref{eq-near_right_singularity}) but keeping only the
leading order term :
\begin{displaymath}
F(\eta) = a_0(1-\eta)^{3/2} + O((1-\eta)^{5/2}).
\end{displaymath}
In terms of $\w$ and $t$,
\begin{displaymath}
n(\w,t) = a_0\ws(t)^{-x-3/2}(\ws(t)-\w)^{3/2} +
O((\ws(t)-\w)^{5/2}).
\end{displaymath}
We substitute this expression for $n_\w(t)$ into equations
(\ref{eq-DKE_E}) and (\ref{eq-DKE_P}) and keep only the leading
power of $\w^*(t)-\w$. For the energy, we obtain
\begin{eqnarray}
\label{eq-tip_energy} \nonumber E(\w,t) &=& \Omega_d\ a_0\
\ \ws(t)^{-x-3/2}\ \w^{d/\alpha}\ \left(\ws(t)-\w\right)^{3/2}\\
& &+ O((\ws(t)-\w)^{5/2}).
\end{eqnarray}
For the flux we obtain
\begin{eqnarray}
\label{eq-eq-tip_flux} \nonumber P(\w,t) &=& \frac{75}{8}\
a_0^3\ \w^{3x_0+3}\,\ws(t)^{-3x-9/2}\ \left(\ws(t)-\w\right)^{3/2} \\
& & + O((\ws(t)-\w)^{5/2}).
\end{eqnarray}
In both these expressions we can write powers of $\w$ as $\w^y =
(\ws(t)-(\ws(t)-\w))^y$ and perform a Taylor expansion in
$(\ws(t)-\w)$. Since we are keeping only terms to leading order in
$(\ws(t)-\w)$, we can consistently replace $\w^y$ with $\ws(t)^y$
to yield the following expressions:
\begin{eqnarray}
\label{eq-tip_energy2} \nonumber E(\w,t) &=& \Omega_d\ a_0\
\ \ws(t)^{-x-3/2+d/\alpha}\,\left(\ws(t)-\w\right)^{3/2}\\
& &+ O((\ws(t)-\w)^{5/2}).
\end{eqnarray}
\begin{eqnarray}
\label{eq-eq-tip_flux2} \nonumber P(\w,t) &=& \frac{75}{8}\
a_0^3\,\ws(t)^{3(x_0-x)-3/2}\,\left(\ws(t)-\w\right)^{3/2} \\
& & + O((\ws(t)-\w)^{5/2}).
\end{eqnarray}
We now substitute these expressions into
(\ref{eq-energy_balance2}) and perform the integrals to leading
order. Let us first consider the LHS:
\begin{eqnarray}
\nonumber \int_{\ws(t_1)}^{\ws(t_2)} E(\w,t_2)d\w&=& \Omega_d\ a_0
\int_{\ws(t_1)}^{\ws(t_2)} \ws(t_2)^{-x-3/2+d/\alpha}\
\left(\ws(t_2)-\w\right)^{3/2} d\w + \mbox{h.o.t.}\\
\label{eq-int_E} &=&\frac{2}{5} \Omega_d
\ws(t_2)^{-x-3/2+d/\alpha}\left(\ws(t_2)-\ws(t_1)\right)^{5/2} +
\mbox{h.o.t.}
\end{eqnarray}
Now consider the RHS:
\begin{eqnarray}
\nonumber\int_{t_1}^{t_2} P(\ws(t_1),t)\ dt &=&\frac{75}{8}\
a_0^3\  \int_{t_1}^{t_2}\ws(t)^{3(x_0-x)-3/2}\
\left(\ws(t)-\ws(t_1)\right)^{3/2}dt +\mbox{h.o.t.}
\end{eqnarray}
Let us change integration variables $t\to\ws(t)$:
\begin{eqnarray}
\nonumber&=&\frac{75}{8}\ a_0^3\
\int_{\ws(t_1)}^{\ws(t_2)}\ws(t)^{3(x_0-x)-3/2}\
\left(\ws(t)-\ws(t_1)\right)^{3/2}\left(\dd{\ws}{t}\right)^{-1}d\ws
+\mbox{h.o.t.}
\end{eqnarray}
We now use the self-similarity condition, (\ref{eq-ss_cond}), to
express the derivative in terms of $\ws(t)$ and perform the Taylor
expansion trick again: for some power, $y$, we write : $\ws(t)^y =
(\ws(t_1)+(\ws(t)-\ws(t_1)))^y = \ws(t_1)^y + \mbox{h.o.t.}$ This
gives
\begin{eqnarray}
 \nonumber &=&\frac{75}{8}\ a_0^3
\ws(t_1)^{3(x_0-x)-3/2}\,\ws(t_1)^{-(3x_0-2x-d/\alpha)}\,
\int_{\ws(t_1)}^{\ws(t_2)}\left(\ws-\w\right)^{3/2} d\ws +\mbox{h.o.t.}\\
\nonumber &=&\frac{75}{8}\ a_0^3 \ws(t_1)^{3(x_0-x)-3/2}\
\left(\dd{\ws}{t_1}\right)^{-1}
\frac{2}{5}\left(\ws(t_2)-\ws(t_1)\right)^{5/2} + \mbox{h.o.t.}\\
\label{eq-int_P}
\end{eqnarray}
Equating (\ref{eq-int_E}) and (\ref{eq-int_P}) and letting $t_1
\to t_2$, we obtain energy balance criterion
\begin{equation}
\label{eq-ss_again} \dd{\ws}{t} \ \ws^{2x-3x_0+d/\alpha} = 1
\end{equation}
which is equivalent to the self-similarity condition
that (\ref{eq-ss_cond}) be time independent.

Now suppose we take $t_2-t_1$ to be small but finite and allow
$t_2\to\ts$. It is clear that the energy flux, (\ref{eq-int_P}),
entering the region $[\ws(t_1),\ws(t_2)]$ in this last increment
of time before the front reaches infinity is either infinite, a
finite quantity, or zero.  Our hypothesis is that the flux is
finite. It certainly cannot be infinite since the entire system
has finite energy. It seems unreasonable that it should be zero
since the front presumably requires a supply of energy to continue
moving. A finite value for the integrated flux requires that the
power of $\w(t_2)$ in (\ref{eq-int_E}) is zero. This hypothesis
leads to a unique value for the slope, which we shall denote by
$x_c$. From (\ref{eq-int_E}):
%\begin{eqnarray}
%\nonumber \ws^{3(x_0-x)-\frac{3}{2}}\
%\left(\dd{\ws}{t}\right)^{3/2} &=&
%\ws^{3(x_0-x)-\frac{3}{2}-\frac{3}{2}(2x-3x_0+\frac{d}{\alpha})}\\
%\label{eq-marginal}&=&\ws^{-6\left(x-x_0-\frac{2\gamma-3\alpha}{12\alpha}\right)}.
%\end{eqnarray}
\begin{eqnarray}
\nonumber & & \frac{2}{5} \Omega_d
\ws(t_2)^{-x-3/2+d/\alpha}\left(\ws(t_2)-\ws(t_1)\right)^{5/2} +
\mbox{h.o.t.}\\
&=&\Omega_d \ws(t_2)^{-x-3/2+d/\alpha} \left(\dd{\ws}{t_2}\right)
^{\frac{5}{2}}\ \frac{2}{5}(t_2-t_1)^\frac{5}{2} +\mbox{h.o.t.}\\
\label{eq-marginal}&=&\ws(t_2)^{-6\left(x-x_0-\frac{2\gamma-3\alpha}{12\alpha}\right)}\
\Omega_d \frac{2}{5}(t_2-t_1)^\frac{5}{2} +\mbox{h.o.t.}.
\end{eqnarray}
In order that this remain finite but non-zero as $\ws(t_2)\to
\infty$, we require
\begin{equation}
x=x_c=x_0+\frac{2\gamma-3\alpha}{12\alpha}
\end{equation}
The physical intuition behind this hypothesis is the following. If
$x>x_c$, the front speed is slower than the critical value. The
power in (\ref{eq-marginal}) is positive and the energy in the tip
diverges as $\ws(t_2)\to\infty$. The front is moving too slowly
for the amount of flux flowing into it so that energy begins to
pile up at the tip. On the other hand, if $x<x_c$, the front moves
faster than the critical value. Then energy in the tip decays to
zero as $\ws(t_2)\to\infty$ which means that the front is moving
too fast for the amount of energy supplied to it. Physically we
expect that the former situation would tend to speed up the front
and the latter would tend to slow it down thus providing the
system with a self-regulatory mechanism which selects the marginal
slope, $x_c$.

%Conclusions and open questions
\section{Concluding Remarks} \label{sec-conclude}

From the work presented here, we are beginning to get a clearer
understanding of how the K-Z spectrum is set up in this model and
the origins of the anomalous spectrum in the wake. The most
obvious and important question which we have not addressed at all
here is that of whether any of this analysis is relevant for the
full kinetic equation, (\ref{eq-4WKE_k}). It seems unlikely that
the detailed structure of the nonlinear front would carry over to
the integral version of the kinetic equation. Indeed, it is
difficult to see how a sharp front tip could co-exist with the
k-space integrations on the RHS of (\ref{eq-4WKE_k}).

Nevertheless, anomalous behaviour has been observed
\cite{galtier2000} in numerical simulations of the full three-wave
kinetic equation as already mentioned. This gives us reason to
hope that some of the qualitative ideas contained here can be
extended to more general kinetic equations. In particular, if the
dynamics of the front is indeed regulated by a critical speed
hypothesis of the type proposed in section \ref{sec-margin}, then
there is hope that an analogue can be formulated even in the
absence of a sharp tip.

One might also speculate that this mechanism might occur in the
evolution of the Kolmogorov spectrum of hydrodynamic turbulence
since the direct cascade in this case is also of finite energy
capacity. Unfortunately, in the absence of a closed kinetic
equation for hydrodynamic turbulence it is not clear how one would
even begin to address this question from a mathematical point of
view. Nevertheless, for the purpose of stimulating debate, let us
make the following conjecture. In far from equilibrium situations
with stationary states which have finite capacity, the evolution
towards the stationary spectrum takes place in two stages. In the
first stage, $0<t<\ts$, which is rapid, the system attempts to
close the connection with the dissipative sink at very high (or
very low) wave-numbers. In that stage, entropy production is
positive as the system attempts to explore all the available phase
space subject to the constraint of energy conservation. The wake
spectrum is steeper (shallower if the sink is at $\k=0$) than that
of the final stationary state. This slope is determined by the
requirement that a finite amount of energy per unit time is
delivered to the front tip at all times less that $\ts$. After
$t=\ts$, energy is no longer conserved but entropy production is
still positive as the system now explores a larger volume of phase
space. However, the entropy production now decreases as a new
front with the K-Z spectrum in its wake (between the front and
$\k=\infty$) moves towards lower wave-numbers and invades the
steeper spectrum set up during the first stage of evolution. While
it may be difficult to confirm these conjectures in a quantitative
manner for the variety of situations to which we suggest these
ideas apply, it should not be too difficult to establish them (or
prove them incorrect) qualitatively.

\section*{Acknowledgements}
The authors would like to thank Oleg Zaboronski and Sergey
Nazarenko for numerous helpful discussions. CC and ACN would like
to acknowledge the hospitality of the Erwin Schrodinger Institute
during the 2002 Program on Developed Turbulence.  We are grateful
for financial support from NSF Grant 0072803, the EPSRC and the
University of Warwick.

\section*{Note Added in Proof}
We have subsequently analysed a second order model equation whose
structure is similar to that of the differential kinetic equation
studied in this paper but is analytically and numerically more tractable. We have
found qualitatively similar behaviour. However the value of the
anomalous exponent appears to differ from that which would be
predicted by our critical front speed hypothesis.

\bibliographystyle{plain}

\bibliography{main}
%Analysis and numerics for the self-similar ode

\appendix
\section{Analysis of the Self-similarity Equation and the Critical
Front Speed} \label{sec-ode}

In this appendix we shall analyse the similarity equation
(\ref{eq-ss_dke}) further. We wish to check that it does indeed
admit solutions which reproduce the critical behaviour which we
have ascribed to the solutions of its antecedent PDE in section
\ref{sec-margin}. Specifically, we are interested in solutions
which match the $F(\eta) \sim (1-\eta)^{3/2}$ singularity at
$\eta=1$ to a power law solution, $F(\eta) \sim \eta^{-x}$ as
$\eta\to 0$.

Let us write (\ref{eq-ss_dke}) in the form
\begin{equation}
\eta^{-1+d/\alpha}\left(xF+\eta\dd{F}{\eta}\right)=
(2(x-x_0)-\kappa_d)\ \ddd{ }{\eta}\left(\eta^sF^4\ddd{
}{\eta}\left(\frac{1}{F}\right)\right),
\end{equation}
and look for solutions of the form $F(\eta)=A\eta^{-y}$.
Substituting this form we obtain
\begin{eqnarray}
\nonumber(-y+x)\ \eta^{\frac{d}{\alpha}-1-y} &=&
3A^2y(y-1)(x_0-y)(3x_0-3y-1)\\
\label{eq-ss_dke_2} & &(2(x-x_0)-\kappa_d)\ \eta^{3 x_0-3y-2}
\end{eqnarray}
We see that we can have the following solutions $y=x$ where $x$
takes one of the following values,
\begin{eqnarray}
 x&=&0\\
 x&=& 1\\
 x&=& x_0\\
 x&=& x_0-1/3\\
\end{eqnarray}
The first pair are the thermodynamic spectra, the second pair are
the K-Z energy and particle spectra respectively. A fifth special
value of $y$ is
\begin{equation}
\label{eq-singular_x}x= x_0+\kappa_d/2.
\end{equation}
What we observe from numerical solution of the O.D.E.
(\ref{eq-ss_dke_2}), described below, is as follows. If we choose
a value of $x$ which is not $x_c=x_0+\kappa_d/4$, then the
solution, $\eta^{-x}$, makes a transition near $\eta=0$ to the
state $\eta^{-y}$, where $y=x_0+\kappa_d/2$. This behaviour near
$\eta=0$ balances the leading order divergences on both sides of
equation (\ref{eq-ss_dke_2}) as $\eta\to 0$. It might be noted,
although it may have no relevance to the problem under
consideration, that this value for the exponent, $y$, leads to
front dynamics, $\ws(t) = (\ts-t)^b$ with zero $b$. This spectrum
was never observed in the solutions of the P.D.E.
%
%the final spectrum is a singular solution of the equation since
%the coefficient on the LHS of (\ref{eq-ss_dke}) diverges for this
%value of x. For example $y \to x_0+\kappa_d/2$ as $\eta \to 0$. It
%turns out that the final form, (\ref{eq-singular_x}), which is the
%only one of the five spectra to balance the order of the
%divergences on both sides of equation (\ref{eq-ss_dke_2}) as $\eta
%\to 0$, is the generic singularity of this equation near $\eta=0$.
%However it is {\it not} the spectrum we expect to see physically.
%From (\ref{eq-b}) we see that the front speed is infinite for this
%choice of $x$.

To  find solutions of (\ref{eq-ss_dke}) which do not exhibit the
$x_0+\kappa_d/2$ scaling as $\eta\to0$,  we decided to perform a
set of numerical experiments. Let us write out the RHS explicitly
so that we can see exactly the equation which we wish to solve:
\begin{eqnarray}
\nonumber\frac{1}{2(x-x_0)-\kappa_d}\
\eta^{-1+\frac{d}{\alpha}}& &\left(xF+\eta\dd{F}{\eta}\right)= -\eta^sF^2\ddddd{F}{\eta} - 2s\eta^{s-1}F^2\dddd{F}{\eta}\\
\label{eq-ss_dke_3}& &+2\eta^sF\left(\ddd{F}{\eta}\right)^2 +
8\eta^s\left(\dd{F}{\eta}\right)\ddd{F}{\eta}\\
\nonumber& &+4s\eta^{s-1}\left(\dd{F}{\eta}\right)^3
+4s\eta^{s-1}F\dd{F}{\eta}\ddd{F}{\eta}\\
\nonumber& &+2s(s-1)\eta^{s-2}F\left(\dd{F}{\eta}\right)
-s(s-1)\eta^{s-2}F^2\ddd{F}{\eta}
\end{eqnarray}
The problem with integrating (\ref{eq-ss_dke_3}) on a computer is
that for generic initial conditions, the strong power law
dependences of the right hand side on the independent variable
render the numerics very susceptible to round-off error and
numerical instability. For example, to study the system with
$\gamma=7/2$, $\alpha=1/2$, $d=1$, to which we gave a lot of
consideration in section \ref{sec-pde_num} due to its relatively
large anomaly, we are required to take $3x_0+2=22$.

To get around this difficulty, and to aid visualisation of the
global properties of the equation, we make the following change of
variables :
\begin{eqnarray}
\nonumber F(\eta) &=& \eta^af(\tau)\\
\label{eq-autonomise} \dd{F}{\eta}&=& \eta^{a-1}g(\tau)\\
\nonumber \ddd{F}{\eta}&=& \eta^{a-2}h(\tau)\\
\nonumber \dddd{F}{\eta}&=& \eta^{a-3}k(\tau),
\end{eqnarray}
where $\tau=\log(\eta)$. By choosing
\begin{equation}
\label{eq-a} a=\frac{1}{2}\left( \frac{d}{\alpha} -s +3 \right) =
-x_0-\frac{\kappa_d}{2} ,
\end{equation}
we can cancel all the power dependence from the system and
eventually recast (\ref{eq-ss_dke_3}) as the the following
autonomous fourth order system:
\begin{eqnarray}
\nonumber \dd{f}{\tau} &=& g-af\\
\nonumber \dd{g}{\tau} &=& h-(a-1)g\\
\label{eq-auto_system} \dd{h}{\tau} &=& k-(a-2)h\\
\nonumber f^2\dd{k}{\tau} &=& -\frac{1}{2(x-x_0)-\kappa_d}\ \left( g+xf\right) -(a+2s-3)f^2k + 2fh^2\\
\nonumber & &+8g^2h + 4sg^3 + 4sfgh +2s(s-1)fg^2 -s(s-1)f^2h.
\end{eqnarray}
Only the region, $f\geq 0$ makes physical sense since $n_\w$, and
hence $F$, cannot be negative. This system is singular on the
hyperplane $f=0$. This system is much easier to integrate
numerically and has the added advantage that we can determine the
presence of fixed points which are not obvious in the original
differential equation. Let us determine these points. It is
obvious that the RHS of (\ref{eq-auto_system}) has a trivial zero
at $(f,g,h,k)=(0,0,0,0) = O$ but this is clearly a  singular point
due to the factor of $f^2$ on the LHS. A second pair of nontrivial
(and nonsingular) fixed points can be shown after quite a bit of
algebra to exist at the points $P_{\pm}=(f_0, af_0, a (a-1)f_0,
a(a-1)(a-2)f_0)$, where
\begin{equation}
f_0=\pm\left(18a(a+1)(a+(s-2)/3)(a+(s-3)/3)\right)^{-\frac{1}{2}}.
\end{equation}
We are naturally only interested in the point $P_+$ since $P_-$
lies in the negative $f$ region. The factors $(a+(s-2)/3)$ and $
(a+(s-3)/3)$ are interesting for the following reason. If we
substitute back in $s=3x_0+2$ and the value of $a$ from
(\ref{eq-a}), these two factors are simply $-\kappa_d/2$ and
$-\kappa_i/2$ respectively. At the transition points between
finite and infinite capacity the fixed point runs away to infinity
which suggests that it has a central role to play in organising
the critical solution in the finite capacity case.

Let us now look for a numerical solution of this system which
mirrors the solutions of the differential kinetic equation. All
the simulations in this section were done using the transformed
system, (\ref{eq-auto_system}), for which a standard
out-of-the-box adaptive Runge-Kutta routine seemed to work fine.
Suppose we want a solution for the wake of the form $F(\eta) \sim
\eta^{-x}$ as $\eta\to 0$. In the new variables this is equivalent
to demanding
\begin{eqnarray*}
f(\eta) &\sim& A\eta^{-a-x}\\
g(\eta) &\sim& -x f(\eta)\\
h(\eta) &\sim& x(x+1)f(\eta)\\
k(\eta) &\sim& -x(x+1)(x+2)f(\eta)
\end{eqnarray*}
as $\eta\to 0$ or as $\tau\to \-\infty$. If $x<-a=x_0+\kappa_d/2$,
which is the case, then we observe that the wake is described by a
trajectory for which $(f,g,h,k) \to (0,0,0,0)$ as $\tau\to
-\infty$. The wake is the singular point, $O$, of the system,
(\ref{eq-auto_system}). As we approach $\eta\to 1$ we must
reproduce the front structure described in section \ref{sec-ss}.
Thus we require a trajectory for which $(f,g,h,k) \to F$ as
$\tau\to 0$, where $F= (0,0,\infty,\infty)$. We require a
trajectory which links these two singular points.

In our numerical simulations we rescaled the variables as follows,
\begin{eqnarray*}
f &\to& \frac{f}{f_0}\\
g &\to& \frac{g}{af_0}\\
h &\to& \frac{h}{a(a-1)f_0}\\
k &\to& \frac{k}{a(a-1)(a-2)f_0}
\end{eqnarray*}
This maps the point $P_+ \to (1,1,1,1)$ for ease of visualisation.
The only difference is that since $a<0$, the signs of $g$ and $k$
are swapped so that the wake part of the solution must now
approach $O$ from the direction $(0^+,0^+,0^+,0^+)$ and the front
tip, $F$,  is now at $(0^+, 0^+, \infty, -\infty)$.

It turns out to be difficult to find a trajectory linking $F \to
O$. We performed a series of experiments integrating backwards
from $1-\eta = \varepsilon$, with $\varepsilon << 1$, towards
$\eta=0$. Because we cannot specify initial data exactly at the
singular point, $F$, we were required to manually tune the initial
conditions quite a bit in order to to reproduce the $F(\eta) \sim
(1-\eta)^{3/2}$ structure near the tip. The remaining adjustable
parameter is the equation is $x$. We again choose to study the
case $\gamma=7/2$, $\alpha=1/2$, $d=1$. Two generic types of
trajectory emerge as we vary $x$. We visualise these trajectories
in 4 dimensional phase space as a pair of projections of the
actual trajectory onto the $(f,g)$ and $(h,k)$ planes
respectively, hence the apparent intersections.

\begin{figure}[ht]
\begin{center}
\epsfig{file=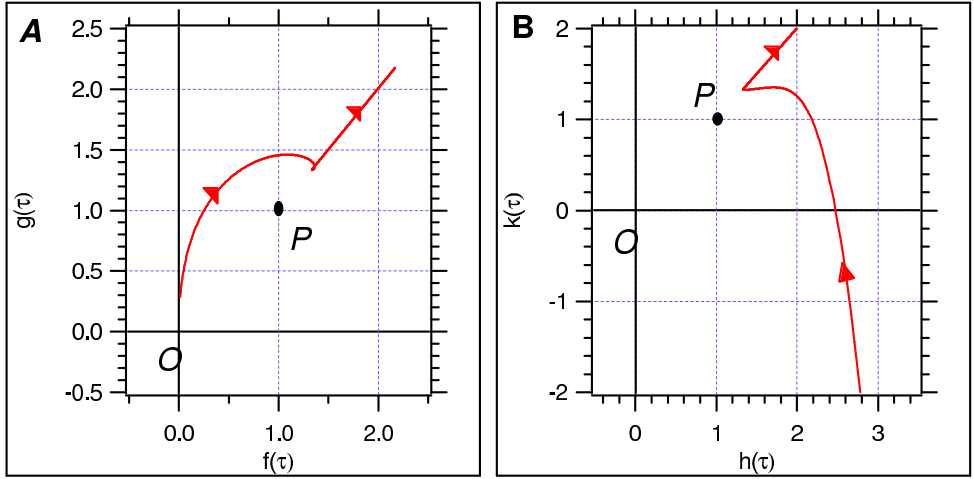,height=2.5in,angle=0}
\end{center}
\caption{\label{fig-traj_xc_plus}Trajectory in $(f,g,h,k)$ space
for $x=7.56275$, which is less than the critical value.}
\end{figure}

\begin{figure}[ht]
\begin{center}
\epsfig{file=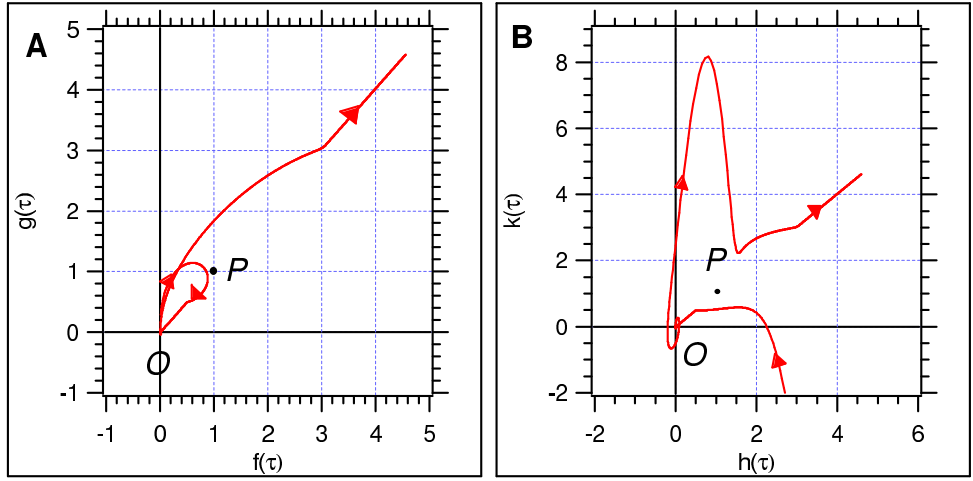,height=2.5in,angle=0}
\end{center}
\caption{\label{fig-traj_xc_minus}Trajectory in $(f,g,h,k)$ space
for $x=7.56975$, which is greater than the critical value.}
\end{figure}

\begin{figure}[ht]
\begin{center}
\epsfig{file=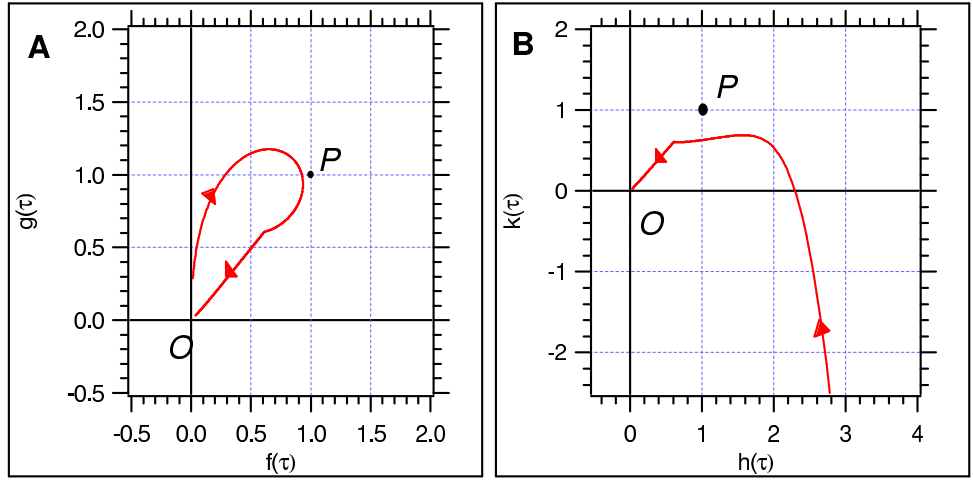,height=2.5in,angle=0}
\end{center}
\caption{\label{fig-traj_xc}Trajectory in $(f,g,h,k)$ space for
$x=7.56875\approx x_c$. Since it is impossible to specify the data
exactly, if we integrate for long enough in $\tau$ this trajectory
eventually deflects away from $O$. }
\end{figure}

\begin{figure}[ht]
\begin{center}
\epsfig{file=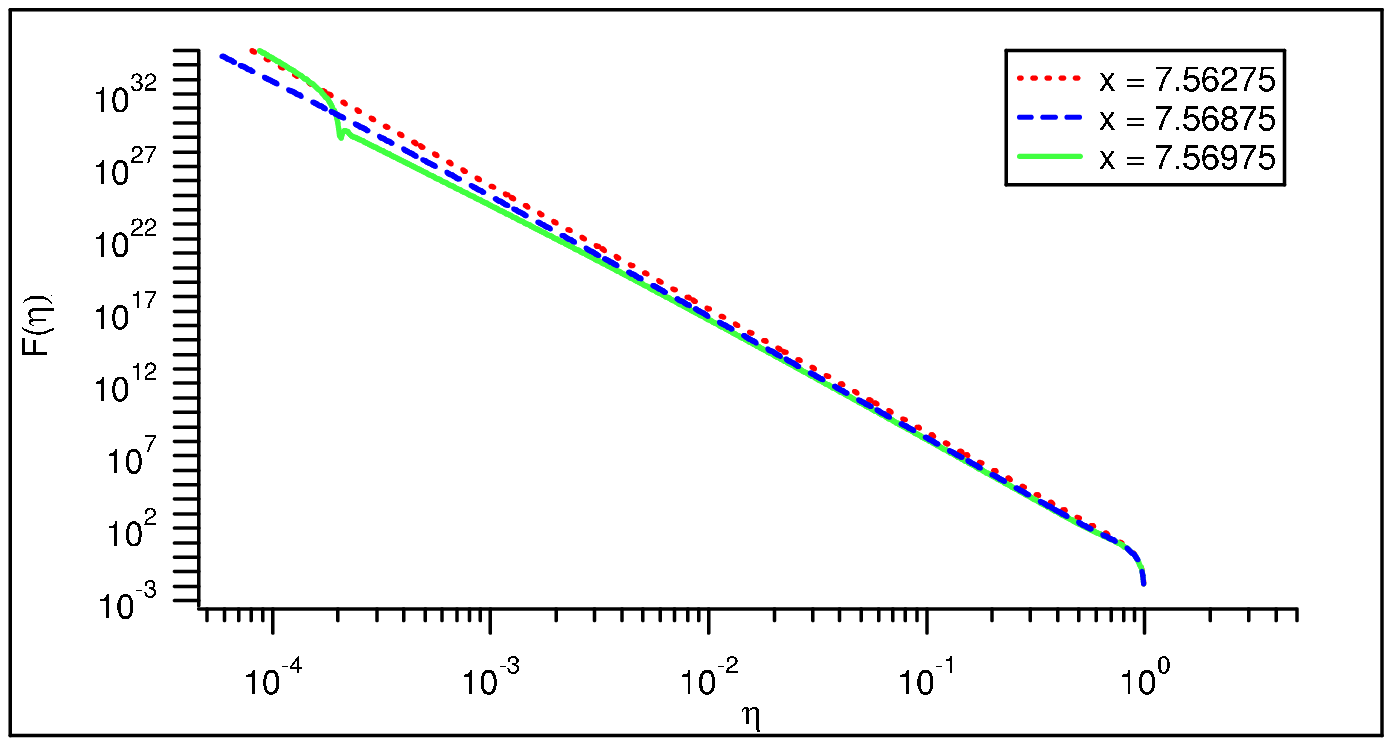,height=2.5in,angle=0}
\end{center}
\caption{\label{fig-asymp_slopes} Plots of the $F(\eta)$
associated with the trajectories in figures
\ref{fig-traj_xc_plus}, \ref{fig-traj_xc_minus} and
\ref{fig-traj_xc} after converting from the $(f,g,h,k)$ and $\tau$
variables back to $F$ and  $\eta$ variables. The scales are
log-log.}
\end{figure}
Figure \ref{fig-traj_xc_plus} shows what happens when $x$ is
slightly less than $x_c$. The trajectory leaves the front tip and
heads towards the fixed point, $P$, but deflects to the right and
is quickly attracted onto the line corresponding to $F(\eta)\sim
\eta^{-x_0-\kappa_d/2}$. Figure \ref{fig-traj_xc_minus} shows the
corresponding trajectory when $x$ is slightly greater than $x_c$.
This time the trajectory deflects to the left as it approaches the
fixed point, $P$, and is attracted towards the singular point,
$O$. Before it reaches there however, it is deflected and is
rapidly attracted back onto the $F(\eta)\sim
\eta^{-x_0-\kappa_d/2}$ solution again. The transition point
between these two trajectories is illustrated in figure
\ref{fig-traj_xc}. Further analysis is required to determine the
exact nature of the critical trajectory. It is not clear whether
it will eventually deflect away from $O$ for sufficiently small
$\eta$. Of course, in practical terms there is a limit placed on
the extent of the scaling region by the left boundary of the
inertial range which cannot extend all the way to 0 in a real
experiment.

The system is very sensitive to the value of $x$. The values of
$x$ for the trajectories shown differ only in the fourth decimal
place. In figure \ref{fig-asymp_slopes}, the form of the function
$F(\eta)$ is shown for the three cases discussed above, after
transforming back from the $(f,g,h,k)$, $\tau$ variables. The
integration was again done for the case $\gamma=7/2$,
$\alpha=1/2$, $d=1$. Despite the small difference in the values of
$x$ in the equation the difference in asymptotic slopes is $7.56$
for the critical slope versus $8.5$ for the other two.

\label{app-ode}

%numerics.tex - outline of the numerical techniques.

\section{Outline of Numerical Methods}
\label{app-pde_num} In order to study how the K-Z spectrum is set
up from some given initial condition we wrote some code to solve
(\ref{eq-dke}) numerically. In this appendix we briefly outline
the approach used.  Write equation (\ref{eq-dke}) in the form
\begin{equation}
\label{eq-DKE3} \pd{n_\w}{t}=f\left[n_\w\right]\pdddd{n_\w}{\w} +
g\left[n_\w\right]\pddd{n_\w}{\w} + h\left[n_\w\right] +
F\left[n_\w\right] - \nu\left[n_\w\right],
\end{equation}
where
\begin{eqnarray}
f\left[n_\w\right]&=& -\Lambda^{-1}_\w \w^sn^2\\
g\left[n_\w\right]&=& -2\Lambda^{-1}_\w s\w^{s-1}n^2\\
\nonumber h\left[n_\w\right]&=& \Lambda^{-1}_\w\left\{\w^s\left(2n\left(\pdd{n}{\w}\right)^2+8\left(\pd{n}{\w}\right)^2\pdd{n}{\w} \right)\right. \\
\nonumber &&\hspace{1.0cm}+2s\w^{s-1}\left(2\left(\pd{n}{\w}\right)^3 + 2n\pd{n}{\w}\pdd{n}{\w} \right)\\
&& \hspace{1.0cm}+s(s-1)\w^{s-2}\left.\left(2n\left(\pd{n}{\w}\right)^2 - n^2\pdd{n}{\w}  \right)\right\}\\
\Lambda_\w&=& \frac{\Omega_\k}{\alpha}\w^{\frac{d}{\alpha}-1}\\
s&=&3x_0+2.
\end{eqnarray}
We are now including forcing and damping terms,
$F\left[n_\w\right]$ and  $\nu\left[n_\w\right]$ which can be
chosen as appropriate. We performed the following implicit time
discretisation,
\begin{eqnarray}
\nonumber\label{eq-DKE4} \frac{n_\w(t+\Delta t)-n_\w(t)}{\Delta
t}&=& f\left[n_\w(t)\right]\pdddd{n_\w(t+\Delta t)}{\w} +
g\left[n_\w(t)\right]\pddd{n_\w(t+\Delta t)}{\w}\\
& & + h\left[n_\w(t)\right] + F\left[n_\w(t+\Delta t)\right]
-\nu\left[n_\w(t + \Delta t)\right],
\end{eqnarray}
with the aim of enhancing the stability of the higher order
derivatives. This can be rearranged to yield a time stepping
algorithm in the form
\begin{equation}
n_\w(t+\Delta t) = L^{-1}\left[n_\w(t)\right]
B\left[n_\w(t)\right],
\end{equation}
where
\begin{eqnarray}
\label{eq-L} L\left[n_\w\right]&=& 1 +\Dt\
f\left[n_\w\right]\pdddd{ }{\w} +
\Dt\ g\left[n_\w\right]\pddd{ }{\w}\\
B\left[n_\w\right]&=&n_\w + \Dt\ h\left[n_\w\right] + \Dt\
F\left[n_\w\right].
\end{eqnarray}
The time evolution operator, $L\left[n_\w\right]$, and source,
$B\left[n_\w\right]$, are approximated using centred difference
representations for the derivatives and a standard linear solver
used to perform the inversion at each time-step. Implementing
consistent boundary conditions is a tricky task. For all of the
simulations presented here, the equation was solved from a
compactly supported initial condition either in a frequency
interval sufficiently large that the solution does not reach the
boundary within the time allotted or with the damping chosen
sufficiently strong to prevent the solution from ever reaching the
boundary.

Most of the simulations presented here are of a freely decaying
initial energy distribution so $F[n_\w] = 0$. The damping was
chosen to be zero over most of the interval but increasing
strongly for $\w<\w_L$ and $\w>\w_R$ to produce regions of strong
dissipation at large and small scales. We typically used
approximately $10$ grid points per unit frequency interval in our
discretisation. The simulations presented here used up to 20000
grid-points. The number of grid points is practically limited by
the fact that one must recompute and invert the time evolution
operator, (\ref{eq-L}), at each time-step. Lacking any reliable
stability criteria for the nonlinear discretisation scheme
described above, the choice of time-step was essentially made by
trial and error. Once a time-step was found for which the
evolution seemed stable, we ran with it. We were aided in choosing
this time-step by monitoring the two conservation laws associated
with the total energy and total particle number. Typically, and
unsurprisingly, the numerics become unstable most easily at the
front tip. This instability is worse for steeper spectra when the
front travels faster. The necessity of reducing the time-step to
resolve this structure as the tip increases in speed also put
practical limits on the time interval we could simulate for.
Retrospectively, we would like to do the numerics using an
adaptive grid which tracks the front. Such a method would probably
yield great increases in efficiency and stability. Unfortunately
at the outset, we did not really appreciate that the code would be
required to tackle this problem. We settled for validating our
computations by checking that our final results remained unchanged
when the time-step was reduced by a factor of two.

\label{app-numerics}
\end{document}